\documentclass[10pt,letterpaper]{article}

\usepackage{ccn}
\usepackage{pslatex}
\usepackage{apacite}
\usepackage{graphicx}

\title{Do Deep Neural Networks Model Nonlinear Compositionality in the Neural Representation of Human-Object Interactions?}
 
\author{{\large \bf Aditi Jha (ee1150504@iitd.ac.in)} \\
  Department of Electrical Engineering, 
  Indian Institute of Technology Delhi\\
  Hauz Khas, New Delhi 110016, India
  \AND {\large \bf Sumeet Agarwal (sumeet@iitd.ac.in)} \\
  Department of Electrical Engineering, 
  Indian Institute of Technology Delhi\\
  Hauz Khas, New Delhi 110016, India}

\begin{document}

\maketitle

\section{Abstract}
{
\bf
Visual scene understanding often requires the processing of human-object interactions. Here we seek to explore if and how well Deep Neural Network (DNN) models capture features similar to the brain's representation of humans, objects, and their interactions. We investigate brain regions which process human-, object-, or interaction-specific information, and establish correspondences between them and DNN features. Our results suggest that we can infer the selectivity of these regions to particular visual stimuli using DNN representations. We also map features from the DNN to the regions, thus linking the DNN representations to those found in specific parts of the visual cortex. In particular, our results suggest that a typical DNN representation contains encoding of compositional information for human-object interactions which goes beyond a linear combination of the encodings for the two components, thus suggesting that DNNs may be able to model this important property of biological vision.
}
\begin{quote}
\small
\textbf{Keywords:} 
visual cognition; deep learning; fMRI; human-object interactions; compositionality
\end{quote}

\section{Introduction}

Visual representations formed by the human brain have been of interest particularly for studying invariance in object representations \cite{DiCarlo2012HowDT, Leyla}. It is known that downstream regions of the visual cortex process high-level visual information and that there are specialized regions for processing object- and human-specific information. Perception of human-object interactions by the brain, though, hadn't been studied much until recent work by \citeA{bald} revealed that the neural representations of interactions are not a linear sum of the human and object representations. In fact, there appear to be certain areas in the brain like the pSTS (posterior Superior Temporal Sulcus) which are highly sensitive specifically to human-object interactions \cite{bald, IsikE9145}. The representation of interaction-specific information in the brain might also be thought of as a kind of {\em visual compositionality}: analogous to compositionality in language, one might say that the meaning of complex visual scenes emerges from the meanings of the individual components plus certain rules of composition.

Deep Neural Nets (DNNs) have been widely used in recent years for a variety of computer vision tasks like object and action recognition \cite{VGGS, Simonyan:2014:TCN:2968826.2968890}. While they have reached human-like accuracy in certain settings, in general there isn't much explicit effort to model biological vision in these networks. However, there has been a lot of work in the past few years attempting to compare DNN representations with those of the brain \cite{cichy2019}. Some recent work has also looked at trying to develop DNNs with explicit compositionality \cite{sto17}.

In this work, we examine if typical DNNs represent humans, objects, and in particular their interactions similarly to the brain. We analyse three brain regions involved in high-level visual processing: the LOC (Lateral Occipital Cortex) which processes object-related information, the EBA (Extrastriate Body Area) which is involved in human pose identification, and the pSTS which also processes human information and is known to be specifically sensitive to human-object interactions. We seek to predict the BOLD (fMRI) responses of these regions to human/object/interaction images, from their DNN representations. Such an approach has been previously used to evaluate the correspondence of DNN layers to brain regions \cite{Guclu} and to model visual representations in the brain \cite{pulkit}.

We look at how well the DNN representations predict the selectivity of the regions for  object, human pose, or interaction stimuli. We probe interaction images to see if our approach also infers the sensitivity of the pSTS to interactions. Additionally, we seek to identify features in the DNN model which are interaction-specific, and which might hence represent compositional information, analogous to the pSTS in the actual visual cortex.

\section{Materials and Methods} 
\subsection{Experimental Data}
The data used have been taken from \citeA{bald}. fMRI data were collected from 12 subjects on showing visual stimuli consisting of three kinds of images: humans, objects, and interactions between humans and objects. Four action categories (working on computer, pulling luggage, pushing cart, and typing on typewriter) were included. The human and object images were extracted from the interaction images by cropping out the relevant part and resizing it.

\subsection{Direct classification using a Deep Neural Network}

\begin{figure}[h!]
\centering
\includegraphics[width=0.4\textwidth]{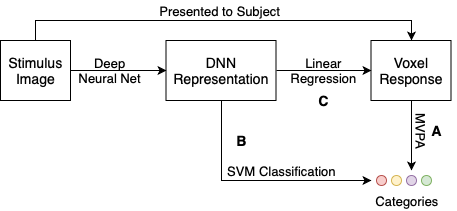}
\caption{Overview of Methodology: Stimulus images were presented to human subjects and voxel responses recorded via fMRI by \protect\citeA{bald}. We pass the same images through a DNN. \textbf{A} \protect\citeA{bald} perform MVPA on the voxel responses to classify the response pattern into one of the four classes. \textbf{B} Direct SVM classification is performed over the final-layer DNN representations of the images. \textbf{C} Linear regression models are trained on the DNN representations to predict voxel responses.}
\end{figure}

We used the pre-trained VGG-16 \cite{VGGS}, a widely-used DNN architecture whose representations have been previously shown to correspond to fMRI recordings from the ventral stream \cite{Guclu}. Three linear SVMs were trained on top of features extracted from this DNN to perform 4-way classification in the three scenarios: humans, objects, and interactions. This is analogous to the Multi-Voxel Pattern Analysis (MVPA) carried out by \citeA{bald}, except for the fact that the SVMs are trained not on voxel activity but DNN representations. The aim was to judge the goodness of the latter for discriminating between action categories in each of the three tasks.

We also looked at whether substructure or specialisation could be identified within the DNN representations, analogous to what we see in subregions of the visual cortex, by quantifying the overlap between the top DNN feature sets picked out by our 3 SVMs. In particular, we looked at those DNN features which were picked using forward selection by the SVM {\em only} for the interaction task, and not for the object or human pose tasks. We sought to examine whether these features capture some nonlinear representation of the interaction, not obtainable by just adding the representations for the object and human constituents separately. This was done analogously to how voxels in the pSTS were analysed by \citeA{bald}.

\subsection{DNN representations to predict voxel responses}

To compare DNN representations with those of the brain, we predict the BOLD response of every voxel to an image using VGG-16's representation for the same image. We average the final-layer representations of 2 or 3 images\footnote{This is because the fMRI recordings have been taken at intervals of every 2--3 images (volume repetition time of functional images is 2s).}, and train a linear regression model on these averaged representations to predict the response of a voxel to the corresponding images. Three such models are trained for each voxel, one per task. The final-layer representations are used because the regions we are looking at process high-level visual information, and a correspondence between such areas and the last convolutional layer of CNNs has been established \cite{Guclu}.

\subsubsection{Region-wise Analysis}
\label{subsec::region-analysis}
Voxels are selected in each region for a particular task based on their Bonferroni-corrected $p$-values for correlation between predicted and actual responses (computed via cross-validation on a training set). For a particular task we then average the correlations (computed on a held-out test set) for all selected voxels for that region. To further ensure that the obtained correlations are not merely flukes, we also trained linear regression models to predict voxel responses from DNN representations of mismatched stimuli, to obtain a baseline correlation level.

We also sought to link the region-wise voxel regression analysis to the task-specific DNN feature sets mentioned above, to check whether DNN features specifically informative for a particular task (say, object recognition) are also specifically more predictive of voxel responses in a particular brain region (say, LOC). If true, this would suggest a correspondence between the substructure or modularity of representation in the DNN and that in the visual cortex. We trained linear regression models for each of the three tasks on only their corresponding task-specific DNN feature sets to predict voxel responses. This allows us to see whether, for instance, the DNN features specific to object recognition also tend to be more predictive of voxel responses (for the same kind of stimuli) in the LOC than in the EBA or pSTS.

\subsubsection{Cross-Decoding Analysis}
Analogous to the MVPA of \citeA{bald} where the classifier trained on interaction images is tested on objects and humans in isolation, we also do a cross-decoding analysis. Every voxel has a linear model trained on the representations of interactions. This is tested on isolated object and human representations (from a held-out test set) to see how well it predicts the voxel's response in those cases. For each region, we obtain the average cross-decoding correlation for those voxels which were selected for the interaction task as above. 

\section{Results}

\subsection{Direct DNN-based classification}

\begin{table}[h!]
\begin{center}
\begin{tabular}{|c|c|}
\hline
{\bf Task} & {\bf Accuracy} \\
\hline
Object classification & 0.90 \\
Human pose classification & 0.74 \\
Interaction classification & 0.86 \\ 
\hline
\end{tabular}
\caption{Direct classification performance of VGG-16 representations on the  \protect\citeA{bald} stimuli.}
\label{directDNN}
\end{center}
\end{table}

The accuracy in each case (Table \ref{directDNN}) is far above chance (0.25), which signifies that the representations formed by VGG-16 are distinguishable into the 4 categories.

\subsubsection{Do the DNN representations also exhibit nonlinear compositionality for human-object interactions?}

We obtained 8 interaction-specific DNN features which do not contribute much to object or human pose classification. Removal of just these features from the all-feature SVMs (Table \ref{directDNN}) decreases interaction classification accuracy (0.71), while object classification accuracy (0.88) and human pose classification accuracy (0.72) do not change much.

These are the features of greatest interest from the point of view of nonlinear compositionality, as they appear to be informative only for the interaction images, and not for classifying just objects or just human poses. \citeA{bald} claimed exactly the same property for the neural encodings recorded from voxels in the pSTS, based on their MVPA analysis: the pSTS shows a comparable accuracy to the LOC and the EBA only in case of human-object interactions, indicating the high sensitivity of the pSTS to such stimuli. Interestingly, they find that the accuracy of the interaction classifier is significantly reduced when tested on isolated humans or objects or even their pattern averages, implying that the relatively high accuracy of the interaction classifier is not attributable only to the object or human in the image.

\begin{figure}
    \centering
    \includegraphics[width=0.5\textwidth]{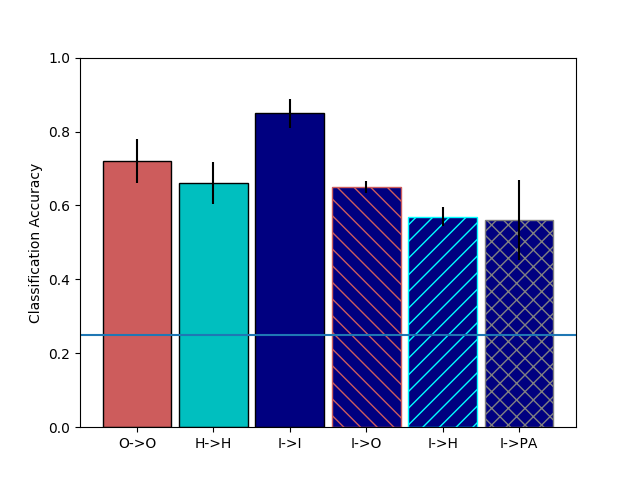}
    \caption{SVM decoding and cross-decoding for interaction-specific DNN features. O: objects; H: human poses; I: human-object interactions; PA: pattern averages of humans and objects. $X\rightarrow Y$ indicates a classifier trained on $X$ and tested on $Y$. Figure inspired by Figure 6 of \protect\citeA{bald}, but using DNN model features rather than voxel responses. The I$\rightarrow$I classifier is significantly more accurate than all others ($p<10^{-13}$ for all pairwise tests, 100 train-test splits).}
    \label{fig:int_dnnf_mvpa}
\end{figure}

To examine if our interaction-specific DNN features indeed exhibit the same behaviour as pSTS neural encodings, we carried out an MVPA-like decoding and cross-decoding analysis on the DNN features, and compared it with the same analysis reported for the pSTS voxels by \citeA{bald}. The results are depicted in Figure \ref{fig:int_dnnf_mvpa}.

The interaction-specific features classify the interaction images with high accuracy ($\sim85\%$). However, the classifier trained on interactions performed much less well on just objects, just human poses, or their pattern averages. Hence, its high accuracy on interactions is not explained solely by human- or object-specific information. Furthermore, even classifiers trained specifically on the object or human pose tasks using these features perform less well on those tasks than what the interaction classifier achieves. These results align well with those of \citeA{bald} for MVPA on the pSTS; hence, these interaction-specific DNN features appear to be analogous to the pSTS in the brain in terms of picking up information beyond just the isolated object or human in a human-object interaction. 

\subsection{DNN representations to predict voxel responses}
\subsubsection{Region-Wise Analysis}

\begin{table}[h!]
\begin{center}
\scalebox{0.85}{
\begin{tabular}{|c|c|c|c|}
\hline
{\bf Region} & {\bf Objects} &{\bf Human Poses}& {\bf Interactions}\\
\hline
LOC & $0.35\pm0.02$ $(13\%)$ & $0.33\pm0.03$ $(12\%)$& $0.2\pm0.02$ $(13\%)$\\
\hline
EBA & $0.33\pm0.03$ $(11\%)$ & $0.39\pm0.04$  $(15\%)$ & $0.25\pm0.03$ $(15\%)$\\
\hline
pSTS & $0.19\pm0.06$ $(9\%)$ & $0.18\pm0.04$ $(12\%)$ & $0.18\pm0.06$ $(8\%)$\\
\hline
\end{tabular}}
\caption{Average Pearson's $r$ and standard deviation for voxel models displaying significant correlation, across 12 subjects (\% of such voxels). The baseline correlation on mismatched stimuli did not exceed 0.06 for any region.}
\label{voxelregp}
\end{center}
\end{table}

Table \ref{voxelregp} shows that the average correlation across all subjects is highest for objects in the LOC and for human poses in the EBA.
This is consistent with the literature since the EBA is known to process human-pose related information, while the LOC processes object information.
The results for the pSTS however indicate that while predictability for interactions is similar to the LOC and EBA, responses to human or object stimuli are predicted relatively much less well in the pSTS. This is consistent with \citeA{bald} who find that the pSTS is less representative of isolated human or object information than the other regions, but similar to them for interactions. 

\subsubsection{Mapping of task-specific DNN features onto brain regions}

\begin{table}[h!]
\begin{center}
\scalebox{0.85}{
\begin{tabular}{|c|c|c|c|}
\hline
{\bf Region} & {\bf Objects} &{\bf Human Poses}& {\bf Interactions}\\
\hline
LOC & $0.27\pm0.02$ $(19\%)$ & $0.09\pm0.03$ $(8\%)$& $0.15\pm0.02$ $(10\%)$\\
\hline
EBA & $0.16\pm0.04$ $(17\%)$ & $0.3\pm0.05$  $(16\%)$ & $0.17\pm0.04$ $(12\%)$\\
\hline
pSTS & $0.19\pm0.04$ $(6\%)$ & $0.12\pm0.03$ $(6\%)$ & $0.13\pm0.04$ $(4\%)$\\
\hline
\end{tabular}}
    \caption{Average Pearson's $r$ and standard deviation for voxel models (trained on only task-specific DNN features) displaying significant correlation, across 12 subjects (\% of such voxels).}
    \label{tab:taskf_region}
\end{center}
\end{table}

Table \ref{tab:taskf_region} shows that the object-specific features are most predictive of the LOC (for object stimuli) while the human-specific features are most predictive of the EBA (for human pose stimuli).
This points towards a correspondence between the task-specific features from the DNN and the two regions which specialize in object and human pose processing.
The pSTS, however, is in general harder to predict as can be seen from the low percentage of significant voxels. However, we again see that while object and human predictability for the pSTS is substantively less than (one of) the other two regions, the interaction-specific DNN features are able to predict interaction responses similarly well in all three regions. 

\subsubsection{Cross-Decoding}

\begin{table}[h!]
\begin{center}
\begin{tabular}{|c|c|c|}
\hline
{\bf Region} & {\bf Objects} & {\bf Human Poses}\\
\hline
LOC & $0.09\pm0.01 $ & $0.05\pm0.01 $\\
\hline
EBA & $0.08\pm0.02 $ & $0.09\pm0.01 $\\
\hline
pSTS & $0.03\pm0.01$ & $0.02\pm0.01$ \\
\hline
\end{tabular}
\caption{Average Pearson's $r$ and standard deviation for cross-decoding for significant voxels (as in last column of Table \ref{voxelregp}) across 12 subjects.}
\label{voxelregcd}
\end{center}
\end{table}

Table~\ref{voxelregcd} shows that the cross-decoding correlations in all three regions are much lower than those for same-task decoding (last column of Table \ref{voxelregp}). Notably, the relative cross-decoding performance is substantially worse for the pSTS than for the other two regions. This is exactly what we would expect, if the pSTS voxels had a specific tendency to encode human-object interaction information which is not obtainable from a linear combination of the constituent human and object segments. The linear models trained on DNN features for the pSTS voxels appear to be learning a different kind of mapping from the other two regions: a mapping which lends itself much less to predicting the responses of the same voxels for just objects or just human poses.

\section{Discussion}
Here we sought to compare DNN representations of human-object interactions with those of the human visual cortex, as a means of modelling the latter computationally. Our results open up the possibility of establishing a correspondence between brain regions and DNN features. In particular, the final-layer DNN features which are found to be useful for action categorisation specifically on the interaction images, are also found to display similar properties to pSTS encodings in terms of capturing information beyond that contained in the object or human subimages.

The region-wise analysis of voxel regression models indicates that DNN representations are predictive of the human brain's responses to visual stimuli, hence implying that the former may model certain aspects of the latter. The cross-decoding analysis reveals that linear models trained to predict pSTS responses on interactions appear to learn a rather different mapping, compared to similar models trained for the LOC or EBA. This suggests, consistent with \citeA{bald}, that the pSTS voxels encode some kind of nonlinear compositionality, and furthermore that our DNN model also contains some such information, which the linear models can pick up when they have been trained to predict pSTS responses in particular. Thus we observe multiple lines of evidence indicative of compositional specialisation of some kind in the DNN representations, analogous to what the pSTS shows for actual neural encodings.

On the whole, this study provides evidence for the supposition that generic final-layer DNN representations of visual stimuli have substructure similar to that found in the visual cortex, and that in particular this seems to include explicit representation for human-object interaction information which goes beyond the individual components. These observations can hopefully motivate an additional direction of study seeking to model biological vision via DNNs, by suggesting that the latter do possess the ability to compositionally represent complex visual scenes as `more than the sum of their parts'.

\section{Acknowledgments}

We are grateful to Chris Baldassano for sharing the stimuli and fMRI data, and to the anonymous reviewers of an earlier submission to CogSci 2019 for helping improve the paper.

\bibliographystyle{apacite}

\setlength{\bibleftmargin}{.125in}
\setlength{\bibindent}{-\bibleftmargin}

\bibliography{main}

\end{document}